\documentclass[10pt,twocolumn]{article}
\usepackage{times}
\usepackage{color}
\usepackage{colortbl}
\usepackage[usenames,dvipsnames,svgames,table]{xcolor}

\usepackage{url}
\usepackage{tikz}
\usepackage{amsmath}
\usepackage{tabularx, booktabs} %% Load packages that you use
\usepackage{balance}
\usepackage{enumitem}
\usepackage{cite}
\usepackage{graphicx}
\usepackage{subcaption}
\usepackage{textcomp}
\usepackage{ifthen}
\usepackage[subtle]{savetrees}

\usepackage{balance}

\usepackage[compact]{titlesec}
\titleformat{\section}
            {\sffamily\Large\bfseries}{\thesection}{1em}{}            
\titleformat{\subsection}
            {\sffamily\large\bfseries}{\thesubsection}{1em}{}            
\titleformat{\subsubsection}
            {\sffamily\normalsize\bfseries}{\thesubsubsection}{1em}{}

\usepackage{MnSymbol}

\ifx\pdfoutput\undefined
\usepackage{graphicx}
\else
\usepackage{graphicx}
\usepackage{epstopdf}
\fi

\usepackage[%
pdftex,              % using ps2pdf vs pdftex
colorlinks=true,     % color the words instead of use a colored box
urlcolor=blue,       % \href{...}{...} external (URL)
filecolor=blue,      % \href{...} local file
linkcolor=black,     % \ref{...} and \pageref{...}
citecolor=black,     % \cite{}
pdftitle={},
pdfauthor={},
pdfsubject={},
pdfkeywords={},
pdfproducer={pdfLaTeX},
pdfpagemode=UseOutlines,  % None, UseThumbs, UseOutlines, FullScreen
bookmarksopen=true   % Does outline start expanded?
]{hyperref}

\usepackage{cleveref}
\crefformat{section}{\S#2#1#3} 
\crefformat{subsection}{\S#2#1#3}
\crefformat{subsubsection}{\S#2#1#3}

\begin{document}
%-------------------------------------------------------------------------------
\date{}

\title{\textbf{\textsf{The workflow motif: a widely-useful performance diagnosis abstraction for distributed applications}}} 

%for single author (just remove % characters)
\author{
{\rm Mania Abdi}\\
Northeastern University
\and
{\rm Peter Desnoyers}\\
Northeastern University
\and
{\rm Mark Crovella}\\
Boston University
\and
{\rm Raja R. Sambasivan}\\
Tufts University
} % end author\maketitle

\maketitle

%-- Use ``fancy'' to include the above Appears in... stuff
\thispagestyle{empty}
%\thispagestyle{fancy}

%-- place any standard commands/environments here to get included in
%-- documents.  When you include this file, you should do it before
%-- the \begin{document} tag.

%%%%%%%%%%%%%%%%%%%%%%%%%%%%%%%%%%%%%%%%%%%%%%%%%%%%%%%%%%%%%%%%%%%%%%
%-- CHANGES:
%-- 07/31/01 -jstrunk- Added command to set the paper margins.

%-- Provides fixed width font for commands and code snips.
\newcommand{\code}[1]{\texttt{\textbf{#1}}}

%-- Terms...  Use this to introduce a term in the paper.
\newcommand{\term}[1]{\emph{#1}}

%-- Provides stylization for e-mail addresses
%\newcommand{\email}[1]{\emph{(#1)}}

%-- Starts a minor section (puts the title inline w/ the text.
\newcommand{\minorsection}[1]{\textbf{#1}:}

%-- Jiri caption
\newcommand{\minicaption}[2]{\caption[#1]{\textbf{#1.} #2}}

%-- Units on numbers: 4KB -> \units{4}{KB}
\newcommand{\units}[2]{#1~#2}

%-- Commands...  i.e. WRITE commands.
\newcommand{\command}[1]{{\sc \MakeLowercase{#1}}}

%-- For notes about things that need to be fixed.
\newcommand{\fix}[1]{\marginpar{\LARGE\ensuremath{\bullet}}
    \MakeUppercase{\textbf{[#1]}}}
%-- For adding inline notes to a draft preceded by your initials
%-- E.g., \fixnote{JJW}{What the heck is a foobar?}
\newcommand{\fixnote}[2]{\marginpar{\LARGE\ensuremath{\bullet}}
    {\textbf{[#1:} \textit{#2\,}\textbf{]}}}

%-- Setting margins: \setmargins{left}{right}{top}{bottom}
\newcommand{\setmargins}[4]{
    % Calculations of top & bottom margins
    \setlength\topmargin{#3}
    \addtolength\topmargin{-.5in}  %-- seems like this should be 1, but .5
                                   %-- balances the text top to bottom
    \addtolength\topmargin{-\headheight}
    \addtolength\topmargin{-\headsep}
    \setlength\textheight{\paperheight}
    \addtolength\textheight{-#3}
    \addtolength\textheight{-#4}

    % Calculations of left & right margins
    \setlength\oddsidemargin{#1}
    \addtolength\oddsidemargin{-1in}
    \setlength\evensidemargin{\oddsidemargin}
    \setlength\textwidth{\paperwidth}
    \addtolength\textwidth{-#1}
    \addtolength\textwidth{-#2}
}

%-- For the tabularx environment... Using L, C, R as the column type
%-- will left, center, or right justify the text.
\newcolumntype{L}{X}
\newcolumntype{C}{>{\centering\arraybackslash}X}
\newcolumntype{R}{>{\raggedleft\arraybackslash}X}

%-- To comment out a swatch of text, use \omitit{blah blah blah}
\long\def\omitit#1{}

%-- Inline title; useful for sub-sub-sections in which you don't want a separate
%-- line for the title.
\newcommand{\inlinesection}[1]{\smallskip\noindent{\textbf{#1.}}}

%-- todo notes

\newenvironment{outlineenv}{\par\color{teal}}{\par}
\newenvironment{pagelenenv}{\par\color{red}}{\par}

\newcommand{\naive}{}% To make sure that \naive isn't already defined    
\def\naive/{na\"{\i}ve}

\begin{abstract}
Diagnosing problems in deployed distributed applications continues to
grow more challenging.  A significant reason is the extreme mismatch
between the powerful abstractions developers have available to build
increasingly complex distributed applications versus the simple ones
engineers have available to diagnose problems in them.  To help, we
present a novel abstraction, the \textit{workflow motif},
instantiations of which represent characteristics of
frequently-repeating patterns within and among request executions.  We
argue that workflow motifs will benefit many diagnosis tasks, formally
define them, and use this definition to identify which
frequent-subgraph-mining algorithms are good starting points for
mining workflow motifs.  We conclude by using an early version of
workflow motifs to suggest performance-optimization points in HDFS.
\end{abstract}

\section{Introduction}
\label{sec:intro}

There is an extreme mismatch between the powerful abstractions that
developers use to build complex, large-scale distributed applications
versus the mostly basic ones that engineers have available to diagnose
problems within deployed instances of them.  We argue that this
mismatch is a key reason why problem diagnosis is extraordinarily
challenging.  Problems can take hours or even days to
diagnose~\cite{Amazon2011Failure, Amazon2017Failure,
Google2018Failure} and over 50\% of engineers' time is spent
debugging~\cite{Odell:2017ei}.

Abstractions to help developers build distributed applications include
(but are not limited to) APIs, allowing complex distributed
applications to be built from simpler ones, load balancers, allowing
applications to (mostly) ignore scaling mechanisms, and datacenter
stack layers (e.g., application, virtualization), which separate
applications from resource sharing.  All these reduce complexity for
developers by allowing complex applications to be build from simpler,
commonly-used \textit{building blocks}, while also \textit{hiding}
implementation details of these building blocks until circumstances
dictate otherwise. 

\begin{figure}[b!]
  %\vspace{-0.3in}
\centering
  \includegraphics[width=3.0in]{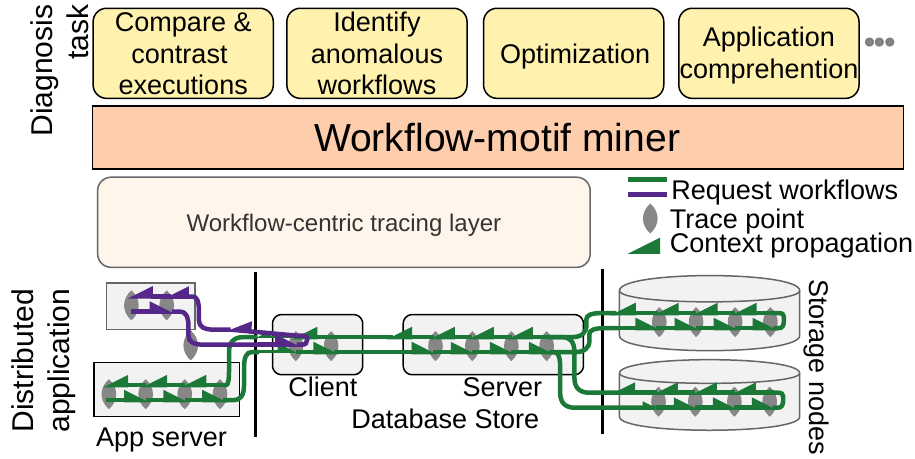}
  %\vspace{-0.15in} 
  \minicaption{Diagnosis tasks, workflow motifs, and
    request workflows}{The top of the figure shows examples of
    diagnosis tasks that involve identifying commonalities within and
    among request workflows.  The bottom shows the workflows of two
    \textsc{READ} requests in a simple distributed application
    comprised of an app server, a database, and a storage system.  The
    first request (purple) hits in the database's client cache; the
    second request (green) requires a storage-node access.}

  \label{fig:workflow}
\end{figure}

In contrast, engineers who must diagnose distributed-application
problems mostly use tools operating on raw, flat logs of
distributed-application activity, with no abstractions
whatsoever~\cite{elasticsearch, Loggly, Nagaraj:2012ur, Xu:2009ui}.
Such tools can struggle to use useful results due to the vast amount
of data they must consider, much of it irrelevant to a given problem.
Tools that use abstractions use ones that leave significant room for
improvement.  They are either too indiscriminate because they focus on
entire requests instead of important behaviors within
them~\cite{Sambasivan:2011vw}, focused only on specifically-marked
regions of requests' execution~\cite{Sigelman:2010uj, OpenTelemetry,
Ding:2015td, Kaldor:2017gp}, or not designed for distributed
applications~\cite{Alimadadi:2018wk}.  Some tools use models of
distributed-application behavior (e.g., culled from logs or
traces)~\cite{Beschastnikh:2014gf, Ohmann:2014cx, Beschastnikh:2011fl,
Zhuang:2014vw}, but such models are static, revealing too much or too
little detail for a specific problem.

We aim to enrich the toolbox of diagnosis abstractions by identifying
an abstraction that is both broadly useful and powerful for a variety
of diagnosis tasks (e.g., the tasks listed at the top of
Figure~\ref{fig:workflow}).  To do so, we observe that many diagnosis
tasks involve identifying commonalities in the workflows of how
requests are processed by a distributed application.  (See the bottom
of Figure~\ref{fig:workflow} for an example of two request workflows
observed in a distributed application.)  For example, the first step
in optimizing the performance of a group of heterogeneous requests,
such as \textsc{READ}s and \textsc{Write}s in a distributed-storage
application, often involves identifying what is common about their
workflows.

We propose the workflow motif, representing frequent
processing patterns observed in requests' workflows.  Examples of such
patterns include known patterns, such as data replication in a
storage application (e.g., Ceph~\cite{Weil:2006ti}), retrieving cached
VMs in a cloud management application (e.g.,
OpenStack~\cite{OpenStack}), or unknown patterns that occur in
unexpected locations, such as problematic background fetches that
synchronize with user-facing main threads~\cite{Gudmundsdottir19}.
These frequent processing patterns represent the \textbf{building
blocks} of distributed applications' runtime behavior.  As such, a
wide range of distributed-application problems can be diagnosed by
identifying them, analyzing them, and understanding their performance
characteristics.

Workflow motifs will mitigate complexity during diagnosis in two ways.
First, they will allow diagnosis tools (and engineers) to analyze
problems in terms of the \textbf{building blocks} of
distributed-applications' runtime behavior.  Second, they will allow
diagnosis tools to easily \textbf{hide} or ignore building blocks that
appear irrelevant to a given problem.  Specifically, they will enable
top-down diagnosis efforts by allowing problematic motifs to be
explored hierarchically.  A motif describing a large processing
pattern may be comprised of many sub-motifs describing simpler
frequent patterns. These sub-motifs will correspond to hierarchical
levels that will be selectively expandable and collapsible.

\minorsection{Key enabler \& approach} The key enabler for workflow
motifs is recent methods for workflow-centric tracing of distributed
applications, which capture graphs (\emph{traces}) of requests'
workflows.  (Please see Sambasivan et al.~\cite{Sambasivan:2016bo} for
a systematization of work in this area.)  Traces are comprised of
records of logging points points executed by requests, linked together
by context (e.g., unique request IDs), which are propagated with
requests.  Sampling techniques are used to keep trace overhead low
enough (e.g., $lt$ 1\%) to be used in production, as is done at many
companies
today~\cite{Sigelman:2010uj,Kaldor:2017gp,Jaeger,OpenTracing}.

% Explicit vs. implicit, isomorphism, traces are graphs
We propose to build on frequent-subgraph-mining algorithms to mine
motifs from traces~\cite{Seno:2003, Nijssen:2004, Lin:2014,
Bhuiyan:2015, Aridhi:2013}. These algorithms are natural starting
points, as they may readily be applied to trace graphs, and they
are able to detect isomorphic patterns corresponding to concurrent or
identical activity, which may comprise a single logical pattern for
top-down diagnosis.
We choose to build on algorithms that mine explicit subgraph
representations instead of implicit ones (e.g.,
graph2vec~\cite{Narayanan:2017:gv} or PCA~\cite{Bishop}) to allow
engineers the ability to understand motifs during their diagnosis
efforts.  However, the motifs we identify could be used as input to
implicit algorithms.  

\minorsection{We present the following contributions} \textit{(1)}
We propose the workflow motif abstraction, representing
characteristics of frequently-repeating patterns within and among
request workflows, and discuss how workflow motifs could
empower a wide range of diagnosis tasks.
\textit{(2)} We present a formal definition of workflow motifs that
allow them to be useful for the range of diagnosis tasks we
consider.  \textit{(3)} We survey existing frequent-subgraph mining
algorithms to identify ones best suited to identifying workflow motifs
and use this information to identify starting points on which to build
motif-mining algorithms.  \textit{(4)} We demonstrate the potential of
workflow motifs for diagnosis by using an initial version of them to
find optimization opportunities in the Hadoop File System~\cite{HDFS}.

\section{Diagnosis tasks}
\label{sec:use_cases}

This section surveys some (but not all) of the diagnosis tasks that
can be informed by workflow motifs.  For these tasks, tools that use
workflow motifs will generate results that more directly identify
problem sources than those that use state-of-the art abstractions:
workflow-centric traces (\textit{trace-based tools}) and
workflow-centric traces that are annotated with spans
(\textit{span-based tools}).  Spans are explicitly marked intervals in
traces that indicate where processing patterns of interest can occur.
Spans can be nested and explored hierarchically.  We expect motifs to
outperform these abstractions because they: \textit{1)} represent
processing patterns that can be observed anywhere in traces, not just
marked regions; \textit{2)} focus on identifying processing patterns
that are common (i.e., frequent) among potentially problematic
requests, which is a key aspect of the diagnosis tasks we
consider; \textit{3)} like spans, enable top-down diagnosis efforts by
allowing processing patterns to be hierarchically explored.

\minorsection{Optimizing performance (existing, mostly performance)}
This diagnosis task involves improving the performance of requests
whose response times fall in different operating regimes of a
distributed-application's performance (e.g., 50-60\textsuperscript{pt}
percentile or the 95-99\textsuperscript{th} percentile).  Though
targeted to performance problems, this diagnosis task may also
involve diagnosing correctness problems that first manifest as
performance issues (e.g., due to retries or
failovers)~\cite{Reynolds:2006tl, guo2013failure}.

\textit{How a motif-based tool could help}: A motif-based tool could
present the slowest motifs observed within requests whose performance
falls within operating regimes of interest.  Examples of slow motifs
in a distributed-application may include: \textsc{read}s of striped
data in a distributed storage-application where I/Os are issued
sequentially instead of concurrently~\cite{Reynolds:2006tl} or
unexpected calls to a slow function in the metadata server when looking up
the location of files in \textsc{read}s and \textsc{write}s.  By
optimizing the processing patterns represented by these slow motifs,
performance of the application as a whole could be improved.

\textit{Contrast to current diagnosis tools}: Today's
trace-based diagnosis tools require developers to examine entire
traces one-by-one (or in groups of requests that execute identical
trace points) to diagnose slow behavior.  They do not identify common
processing patterns within requests that are slow.  Though span-based
tools~\cite{OpenTracing, Jaeger, Sigelman:2010uj} can operate on
processing patterns, they are limited to those that begin and end at
explicitly marked regions (i.e., spans).  As such, they would be
unable identify the unanticipated pattern in the metadata server
(unless the slow function was explicitly bounded by a span).  To our
knowledge, no span-based tool incorporates the concept of frequency
that motifs natively provide.

\minorsection{Contrasting distributed-application executions
  (existing, correctness \& performance)} This diagnosis task involves
contrasting a ``good'' execution 
with a problematic one to identify important differences, as such
differences are promising starting points for diagnosis
efforts.  Differences may manifest in the order of work done to
process identical requests in both executions (i.e., in request
\textit{structures}), or in the
amount of time it takes for the same work to complete
(i.e., just in \textit{latencies}).  Executions could correspond to
instrumentation data from two time periods of a long-running
application (e.g., yesterday vs. today) or two different
regression tests (e.g., before and after a commit).  Problems
identified by this task could be related to performance (e.g., a
function that takes longer to execute in the problematic execution) or
to correctness (e.g., requests that replicate data to only two storage
nodes rather than three in the problematic execution).

\textit{How motif-based tools could help}: A motif-based tool could
identify structural and timing differences between the motifs observed
in the non-problematic and problematic execution.  Doing so would help
users understand how application building blocks have changed between
both executions.  The hierarchical nature of motifs also allows
exploring differences in a top-down manner, first comparing the
highest-level motifs with all sub-motifs collapsed could be compared,
then comparing with the next level of sub-motifs expanded if
additional detail is needed. This hierarchical exploration would
reflect the fact that structural differences in higher-level
functionality, such as changes in services or APIs access, may be be
more important than differences in lower-level functionality, such as
individual functions.

\textit{Contrast to current diagnosis tools}: Spectroscope
~\cite{Sambasivan:2011vw, Sambasivan:2013tq}, the current
state-of-the-art tool for contrasting executions, identifies the
requests that have changed in structure or latencies between two
executions.  It is trace-based, so it leaves the difficult task of
identifying which processing patterns within requests have changed and
whether the same processing pattern is responsible for changes across
multiple heterogeneous requests (e.g., \textsc{read}s
an \textsc{write}s) to engineers.  Both of these activities were
instrumental to diagnosing the problems discussed in Sambasivan et
al.~\cite{Sambasivan:2011vw}.  Existing span-based tools~\cite{Jaeger,
OpenTracing} are limited to patterns in marked regions and are
currently only capable of contrasting individual requests, not entire
executions.

\minorsection{Identifying anomalies (correctness
  \& performance)} This task involves identifying requests
with unusual or rare workflows.  These may represent latent
correctness problems, or ones that first manifest as
performance problems or just performance problems.  Examples of each
category are: a small number of requests that fail silently,
storage-node failures in a distributed-storage application that are
masked by retries to different replicas, or an unusual option
specified in a request,  resulting in execution of a rare,
slow code path.  

\textit{How workflow-motif-based tools could help}: A motif-based tool
could use statistical or machine-learning techniques to learn which
motifs often co-occur in request' workflows and which motifs never
co-occur.  It could present workflow-centric traces of requests whose
workflows violate these learned patterns to engineers.  To help
engineers with their diagnosis efforts, the traces could be annotated
with their constituent motifs.  Those motifs that rarely co-occur and
ones missing their typical co-occurrence partners could be explicitly
marked.

\textit{Contrast to current diagnosis tools}: Existing
trace-based~\cite{Barham:2004uq, Xu:2009ui, Sifter} anomaly detection
tools do not have access to the semantically meaningful, higher-order
information that workflow motifs provide.  As such, we hypothesize
that their predictions may be less accurate and harder to interpret in
certain cases than motif-based anomaly detection tools.  For example,
Sifter~\cite{Las-Casas2019a} identifies rare requests using neural
networks trained on entire traces.  Training them using the motifs
embedded in traces instead may improve their power or accuracy.  We
are not aware of any span-based anomaly detection tools.

\section{Workflow motifs \& mining them}
\label{sec:motifs}

% Formally introduce workflow motifs%
Informally, workflow motifs represent frequent patterns observed in
request workflows augmented with performance characteristics, mined
from workflow-centric traces.  This section defines workflow motifs
formally by identifying basic requirements needed for them to be
broadly useful for the diagnosis tasks discussed in the previous
section (Section~\ref{sec:motifs:defn}).  We use the final motif
definition to determine which existing graph-mining tools are good
starting points on which to build our workflow-motif-mining techniques
(See Section~\ref{sec:progress}).  We start this section by briefly
describing characteristics of workflow-centric traces, a description
that accommodates (but may be more general than) all existing
workflow-centric tracing infrastructures~\cite{Fonseca:2007uc,
Sigelman:2010uj, OpenTracing, OpenTelemetry, Jaeger}.

\begin{figure*}[!t]
    \centering
    \begin{subfigure}{0.68\textwidth}
    	\vspace{-0.03in}
        \includegraphics[height=4.5cm]{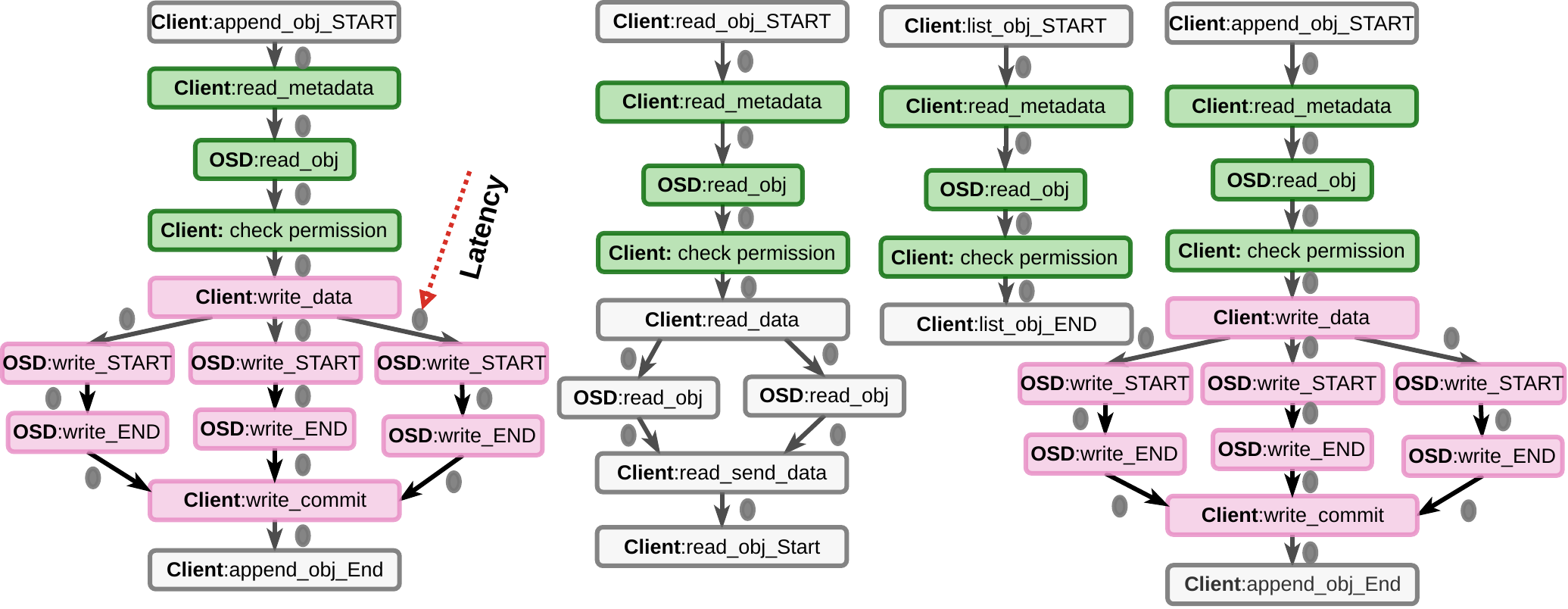}
        \vspace{-0.15in}
        %\caption{Workflow-centric traces}
        \label{fig:design:traces}
        \vspace{-0.05in}
    \end{subfigure}
    \begin{subfigure}{0.28\textwidth}
    	\vspace{-0.2in}
        \centering
        \includegraphics[height=4.5cm]{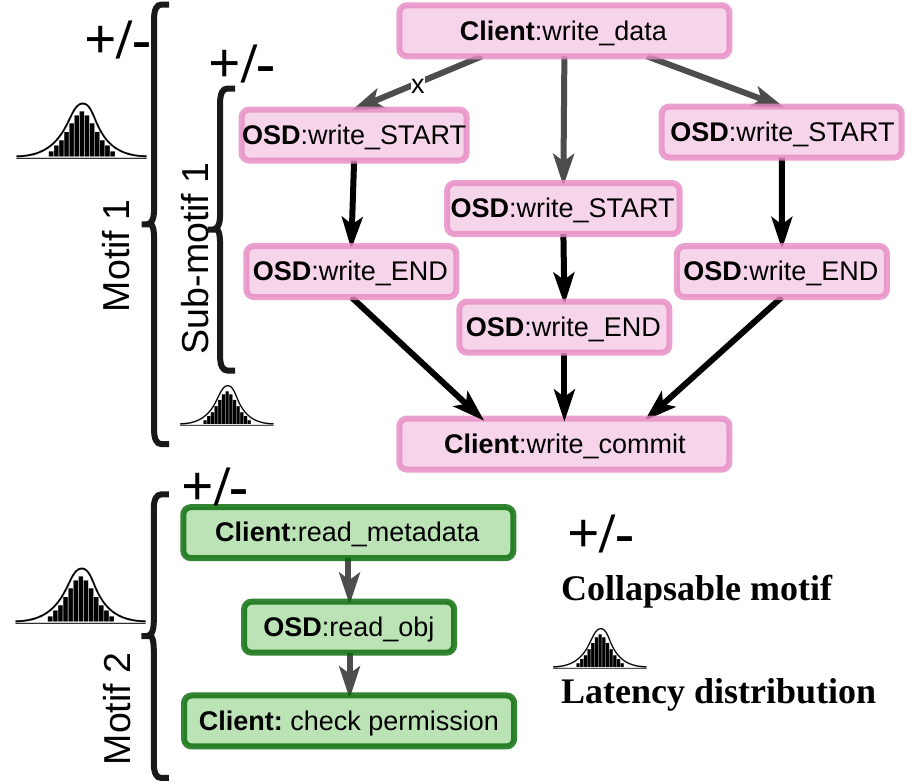}
        \label{fig:design:workflowmotif}
        \vspace{-.15in}
        %\caption{Workflow motifs}
        \vspace{-0.2in}
    \end{subfigure}    
    \vspace{-0.1in}
    \minicaption{Workflow-motif-mining example}
                {The leftmost figure shows a set of
                  hypothetical workflow-centric traces mined from
                  Ceph, a distributed-storage application.
                  The rightmost figure shows the
                  workflow motifs that would be mined from these
                  traces.}
    \label{fig:design:hero}
    \vspace{-0.24in}
\end{figure*}

\minorsection{Workflow-centric traces}: Workflow-centric traces are
directed-acyclic graphs (see Figure~\ref{fig:design:hero}).
Vertices represent trace-point names (e.g., \texttt{OSD:write\_start}
in the figure); edges indicate happens-before relationships and are
labeled with latencies (grey ovals in the figure).  Fan-outs represent
the start of concurrent activity (forks) and fan-ins represent
synchronization (joins).  Trace points are \textit{span markers}
(identified by \texttt{\_START} and \texttt{\_END} in the figure)
bound intervals of semantically-related activity
called \textit{spans}, and \textit{annotations} describe specific
distributed application events.

%-------------------------------------------------------------------------------
% hero diagram
%-------------------------------------------------------------------------------

\subsection{Defining workflow motifs}
\label{sec:motifs:defn}

We start with a common, basic definition of a frequent subgraph since
workflow-centric traces are graphs and frequent-processing patterns
are subgraphs of them.

The standard definition of a frequent subgraph is a subgraph that
appears in a large fraction of input graphs.  Formally, assume a set
of undirected graphs ${\mathcal{G}}=\{G_1, G_2, \ldots, G_k\}$, where
each graph is made up of vertices and edges ${G_i}=({\mathcal{V}_i},
{\mathcal{E}_i})$. Let the set of graphs in which a subgraph $g$
occurs be $\delta_{\mathcal{G}}(g)=\{G_i \,|\, g \subset G_i\,,\, G_i
\in \mathcal{G}\}$.  Then the frequency-based support of $g$ is
$s_{\mathcal{G}}(g)=|\delta_{\mathcal{G}}(g)|/k$. Given a support
threshold $\sigma \in (0, 1)$, the frequent subgraphs contained in
${\mathcal{G}}$ are a set of graphs:

\begin{equation}
  F({\mathcal{G}}) = \{g\;| \;s_{\mathcal{G}}(g) \geq \sigma\}.
  \label{fig:eqn_fg}
\end{equation}

The above definition means that any subgraph of a larger frequent
subgraph is also frequent.

\minorsection{Requirements} We now introduce requirements that
workflow motifs must satisfy in order to be broadly useful.  We use
examples from Ceph~\cite{Weil:2006ti} and HDFS~\cite{HDFS} to motivate
them.

\textit{\underline{R-1}: \textit{Motifs must preserve work
    order}} Some frequent-processing patterns observed in distributed
applications may execute the same work, but differ only in the order
in which they execute it.  Such patterns should be identified as
separate motifs to faithfully represent processing patterns and reveal
problems (e.g., perhaps execution order should be the same).

\textit{Example}: Traces of \textsc{Read} and \textsc{Write}
workflows, we captured by instrumenting Ceph~\cite{Weil:2006ti} show
that they contain processing patterns that differ only in order.
Since objects in Ceph are immutable, \textsc{Write} requests first
check whether the object being written already exists before
performing additional work.  In contrast, \textsc{Read} requests perform
other metadata operations first before executing the existence
check.  Preserving edge order so that these patterns can be identified
as separate motifs would illuminate these cases. 

\textit{\underline{R-2}: \textit{Motifs must be processing
    patterns that are frequent both within traces and across them}}
Certain processing patterns will repeatedly occur many times within a
small set of request workflows, but will not occur in a large set of
them.  Others will occur only a few times within individual workflows,
but occur repeatedly in many different ones.  To avoid missing
important frequent processing patterns (e.g., ones whose performance
could be improved), both types of frequency must be considered when
identifying workflow motifs.

\textit{Example}: Traces \textsc{read} requests in
Ceph~\cite{Weil:2006ti} show a large number of smaller reads from
storage services, with the exact number depending on the size of the
file being read.  These smaller reads do not appear in other requests'
workflows and so not counting within-graph frequency would mean they
would not be identified as motifs.  This may prevent engineers from
learning about them or understanding that they may be important enough
to optimize or mitigate (e.g., by using larger batch sizes when
reading from storage services).

\textit{\underline{R-3}: \textit{Motifs show performance
    characteristics}} For a given motif motif to be useful for
performance diagnosis, it must show the performance characteristics of
the processing patterns it represents.  These performance
characteristics include distributions of the processing pattern's
execution times and edge latencies in the traces it occurs in.
Ideally, execution times should be that of the processing pattern's
critical path in each trace as this is the path on which the pattern's
performance depends.  (Note: for processing patterns that include
forks and joins, the critical path may be different for different
traces in which the pattern is observed.)

For processing patterns that are paths or which incorporate entire
forks and joins, the critical path is defined.  It is simply the
longest path that joins (synchronizes); its latency can be calculated
as the sum of its edge latencies or as the time difference between the
timestamps of its first and last trace points. (The latter method
assumes synchronized clocks or tracing infrastructures that account
for time skew as is common).  For processing patterns that are
fragments of fork/join operations, critical paths are ill-defined if
corresponding forks and joins are not present.  In such cases, we
relax our definition of execution time to be the time difference
between the earliest trace point and the latest timestamp in the
processing pattern.  Processing patterns that appear to be paths may
in fact be fork/join fragments.  To identify such cases, an explicit
flag can be propagated along with request context that indicates
whether a currently executing path is actually a concurrent branch.

\minorsection{Final workflow-motif definition} Building on the
domain-specific requirements above and the definition of frequent
subgraphs (Equation~\ref{fig:eqn_fg}), we define workflow motifs as
follows.  We start from a workflow-centric trace $T$, which is a
weighted directed acyclic graph, where vertices in each trace
$(\mathcal{V})$ represent trace point names and edges represent the
\emph{happens-before}~\cite{Lamport:1978tr} relation between trace
points.  Edge weights indicate latencies between successive trace
points.

\begin{table*}[tb]
    \centering
    \small
          \begin{tabular}{@{}p{12ex}p{39ex}p{9ex}p{15ex}p{20ex}p{10ex}@{}}
      \toprule
                   {}                                                   & {\bf Approach}                              &  \textit{Edge dir (\underline{R-1})}   &    \textit{In/across graphs (\underline{R-2})}   &    \textit{Extra info w/ FSGs (\underline{R-3})}     &                 \textit{Scalability}                \\
          \midrule
                 % Arabesque 4/5, distributed                                      
                   Arabesque~\cite{Teixeira:2015bk}                     &    Join-based                      &     N                               &        Y                             &            Y                     &             $\ast$                             \\
                 % GraMi 2/5, centralized  
                   \textsc{GraMi}~\cite{Elseidy:2014ed}                 &    CSP                                     &     Y                               &        Y                            &            N                              &           $\dagger$            \\
                 % ASAP   1/5, distributed
                  ASAP~\cite{Iyer:2018td}                               &    Join-based \& Sampling          &     N                               &        Y                            &            N                       &           $\ast\dagger$           \\
                 % Gaston 1/5  
                   Gaston~\cite{Nijssen:2004}                           &    Iterative Pattern growth: Paths, trees, graphs                           &     Y                               &        N                               &    N   &      $\dagger$                       \\
                % FSG 0/5, centralized
                   FSG~\cite{Seno:2003}                                 &    Join-based                               &     N                               &        N                            &            N                     &           \\

           \bottomrule
          \end{tabular}
          \vspace{-0.10in} 
          \minicaption{Potential algorithms/systems for workflow motif 
          mining and their properties}{ Listed characteristics are (R-1) 
          edge direction support, (R-2) mining across/within multiple graphs, 
          (R-3) augmented graph information (e.g. latency), scalability 
          via distribution ($\ast$) or optimized algorithm \& implementations ($\dagger$).}
          \label{tbl:graph_mining_survey}
          \vspace{-0.20in}
\end{table*}

Given a set of workflow-centric traces $\mathcal{T} = (T_1, \dots, T_k)$,
let the number of times a
processing pattern (i.e., a subgraph) $p$ occurs within all of
${\mathcal{T}}$ be $\delta_{\mathcal{T}}(p)=\{T_i \;|\; p \subset T_i,
T_i \in \mathcal{T}\}$.
Note that subgraph containment is determined only on structural
properties -- it does not require edge weights to match.
Then
the frequency-based support of $g$ can be defined as
$s_{\mathcal{T}}(g)=|\delta_{\mathcal{T}}(g)|/k$.
Then we define a workflow motifs as a processing patterns that occurs over a
certain number of times in the input traces,
a distribution of its execution times $ExecTimeDist$, and a
distribution of its edge latencies $EdgeLatDist$.
That is, $w$ is a workflow motif of $\mathcal{T}$ if

\begin{eqnarray*}
  w & = & (p,\; ExecTimeDist(p),\; LatDist({\cal E}^{'})) \\
   \mbox{where~~} p & = & ({\cal V}^{'}, {\cal E}^{'}),\\
   \mbox{and}& & s_{\mathcal{T}}(p) > \sigma.
\end{eqnarray*}

Any subgraph of a workflow motif is itself a workflow motif.  Thus
motifs can be hierarchically explored using any meaningful criteria to
mark hierarchical levels (e.g., subgraph size).  To limit the amount
of motif data that must be stored, the criteria could optionally be
pushed into the definition of workflow motifs (e.g., only identify
processing patterns as motifs if they are multiples of n trace points
in size).

The right side of Figure~\ref{fig:design:hero} shows two workflow
motifs mined from the hypothetical Ceph traces traces on the left side
of the figure.  In this mock example, the purple motif happen to be an
entire span and the green one represents behavior in the metadata
server that was not enclosed in a span.  The motifs represent
important \textbf{building block}s of Ceph's run-time behavior
(writing data and reading metadata).  They are annotated with
performance characteristics to help diagnose performance problems.

\section{Initial progress}
\label{sec:progress}

We list five different seminal and state-of-the-art frequent subgraph
mining systems in Table~\ref{tbl:graph_mining_survey}.  Algorithms are
characterized by fundamental approach (i.e. how they expand initial
subgraph candidates), whether they satisfy requirements of edge
direction, ability to mine within and across disjoint graphs, and
support for edge/vertex annotation (i.e. latency/timing information),
and their scalability.  Although most of these algorithms meet some
requirements, none satisfy all.

\minorsection{Building techniques for mining workflow motifs}
We are exploring using Arabesque and Gaston as the starting points on
which to base our workflow-motif-mining techniques.  We have chosen to
explore both concurrently for two reasons: (1) as a large scale
distributed graph mining framework rather than a single algorithm,
Arabesque is both scalable and allows storage of per-subgraph and
per-edge information, such as performance characteristics and (2)
although Gaston is a non-distributed algorithm, its division of graphs
into frequent paths, subtrees, and graphs mimics the structural
characteristics of many of the traces we have seen so far.  We discuss
domain-specific modifications we have made to Arabesque and Grami so
far below.

\emph{\underline{R-1:} (edge directionality):} this is represented via
edge labels; Arabesque and Gaston will restrict subgraph matches to
ones with identically-labeled edges.  For a given edge
${{E}_{ij}}=({{v}_i}, {{v}_j})$, we apply the label $L_{{E}_{ij}}=
L_{{v}_{i}}\ll{32} + L_{{v}_{j}}$.  This distinguishes between
${v}_{i}\rightarrow{{v}_{j}}$ and ${v}_{j}\rightarrow{{v}_{i}}$,
preventing the detection of erroneous subgraphs which do not
correspond to execution order.

\emph{\underline{R-3:} (performance annotation):}
We have extended Gaston to allow annotation of vertices, edges and
subgraph templates; vertices are labeled with the timestamp at which
the tracepoint executed. For subgraph templates, latency distributions
are the differences between the minimum and maximum timestamps
observed in the subgraphs they represent.

\minorsection{Case study}
We describe the potential for workflow motifs to help diagnose
distributed-application problems with a preliminary case study. We
used our modified Arabesque that supports edge direction to find
frequent subgraphs out of traces captured in HDFS.  We annotated these
subgraphs with performance characteristics in an expensive post
processing step to obtain an initial version of workflow motifs.  We
focused on understanding \textit{optimizing} the performance of HDFS
by extracting expensive motifs and understand they they occur.

\emph{Experimental Setup:} 
We feed a set of traces collected from pre-instrumented Hadoop
2.7.2~\cite{BrownTracing} and X-Trace~\cite{Fonseca:2007} running on a
Hadoop cluster with 1 name-node and 3 data-nodes. We create a
benchmark consist of combination of 100 read, 100 write, 100 list and
100 remove of objects. With a support level of 75\%, we were able to
identify 40 motifs.

\emph{Optimization.1 (slow read performance):} To identify motifs corresponding
to slow reads, we selected the read trace motif with maximum execution
time.  This corresponded to a pair of threads, one on the client and
one on the data node, with the data node thread reading an HDFS block
in small 64KB chunks and then writing each chunk synchronously to the
network before beginning the next one, resulting in low disk and
network utilization.  When brought to the Hadoop community, this was
described as an early design decision due to the lack of asynchronous
I/O in Java.  More recent versions have moved this logic to a thread
pool and used \texttt{fadvise(2)} to enable file readahead.

\section{Open questions}
\label{sec:discussion}

We believe the workflow motif will be a powerful abstraction for
diagnosis.  But many questions remain before their potential can be
realized.  We survey some here.  \textit{First}, mining motifs can be
very expensive (exponential time in the worst case).  What other
domain-specific simplifications or heuristics could be used to reduce
mining overhead?  \textit{Second}, mining sparely-instrumented
distributed applications for motifs will yield a few meaningless ones
and mining overly-instrumented systems may result in not enough
patterns being frequent enough to be identified as motifs.  Is it
possible to automatically determine what level of instrumentation
is optimal for workflow motifs, perhaps leveraging current research
into automatically choosing instrumentation~\cite{Ates:2019th,
Zhao:2017co}?  \textit{Third}, it possible to create motif-mining
algorithms that operate in streaming fashion so that motifs can be
used for online diagnosis tasks?

\section{Summary}
\label{sec:conclusion}

There is a mismatch between the abstractions available to help
developers build distributed applications versus those available to
help engineers debug problems in them.  We propose a novel diagnosis
abstraction called the workflow motif, introduce requirements
workflow-motif-mining techniques must meet, and discuss initial
steps towards building them.

{
\renewcommand{\baselinestretch}{1.0}
%\small
\balance
\bibliographystyle{acm}
\bibliography{leitmotif_arxiv_submit}

\begin{thebibliography}{10}

\bibitem{Alimadadi:2018wk}
{\sc Alimadadi, S., Mesbah, A., and Pattabiraman, K.}
\newblock {Inferring Hierarchical Motifs from Execution Traces}.
\newblock In {\em ICSE '18: Proceedings of the 40th International Conference on
  Software Engineering\/} (May 2018).

\bibitem{Aridhi:2013}
{\sc Aridhi, S.}
\newblock {\em Distributed frequent subgraph mining in the cloud}.
\newblock PhD thesis, Universit{\'e} Blaise Pascal-Clermont-Ferrand II, 2013.

\bibitem{Ates:2019th}
{\sc Ates, E., Sturmann, L., Toslali, M., Krieger, O., Coskun, A.~K., and
  Sambasivan, R.~R.}
\newblock An automated, cross-layer instrumentation framework for diagnosing
  performance problems in distributed applications.
\newblock In {\em SoCC '19: Proceedings of the Tenth Symposium on Cloud
  Computing\/} (2019).

\bibitem{Barham:2004uq}
{\sc Barham, P., Donnelly, A., Isaacs, R., and Mortier, R.}
\newblock {Using Magpie for request extraction and workload modelling}.
\newblock In {\em OSDI '04: Proceedings of the 6\textsuperscript{th} USENIX
  Symposium on Operating Systems Design and Implementation\/} (2004).

\bibitem{Beschastnikh:2014gf}
{\sc Beschastnikh, I., Brun, Y., Ernst, M.~D., and Krishnamurthy, A.}
\newblock Inferring models of concurrent systems from logs of their behavior
  with csight.
\newblock In {\em ICSE '14: Proceedings of the 36th International Conference on
  Software Engineering\/} (2014).

\bibitem{Beschastnikh:2011fl}
{\sc Beschastnikh, I., Brun, Y., Schneider, S., Sloan, M., and Ernst, M.~D.}
\newblock Leveraging existing instrumentation to automatically infer
  invariant-constrained models.
\newblock In {\em ESEC/FSE '11: Proceedings of the 19th ACM SIGSOFT Symposium
  and the 13th European Conference on Foundations of Software Engineering\/}
  (Sept. 2011).

\bibitem{Bhuiyan:2015}
{\sc Bhuiyan, M.~A., and Hasan, M.~A.}
\newblock An iterative mapreduce based frequent subgraph mining algorithm.
\newblock {\em IEEE Transactions on Knowledge and Data Engineering 27}, 3
  (March 2015), 608--620.

\bibitem{Bishop}
{\sc Bishop, C.~M.}
\newblock {\em Pattern Recognition and Machine Learning (Information Science
  and Statistics)}.
\newblock Springer-Verlag, Berlin, Heidelberg, 2006.

\bibitem{BrownTracing}
Browntracingframework.
\newblock
  \url{http://brownsys.github.io/tracing-framework/docs/tutorials/gettingstarted.html}.

\bibitem{Ding:2015td}
{\sc Ding, R., Zhou, H., Lou, J.-G., Zhang, H., Lin, Q., Fu, Q., Zhang, D., and
  Xie, T.}
\newblock Log2: {A} cost-aware logging mechanism for performance diagnosis.
\newblock In {\em ATC '15: Proceedings of the 2015 USENIX Annual Technical
  Conference\/} (2015).

\bibitem{elasticsearch}
{Elastic Search}.
\newblock \url{https://www.elastic.co/products/elasticsearch}.

\bibitem{Elseidy:2014ed}
{\sc Elseidy, M., Abdelhamid, E., Skiadopoulos, S., and Kalnis, P.}
\newblock {GraMi}: frequent subgraph and pattern mining in a single large
  graph.
\newblock {\em Proceedings of the VLDB Endowment 7}, 7 (Mar. 2014), 517--528.

\bibitem{Fonseca:2007uc}
{\sc Fonseca, R., Porter, G., Katz, R.~H., Shenker, S., and Stoica, I.}
\newblock {X-Trace: a pervasive network tracing framework}.
\newblock In {\em NSDI '07: Proceedings of the 4\textsuperscript{th} USENIX
  Symposium on Networked Systems Design and Implementation\/} (2007).

\bibitem{Fonseca:2007}
{\sc Fonseca, R., Porter, G., Katz, R.~H., Shenker, S., and Stoica, I.}
\newblock X-trace: A pervasive network tracing framework.
\newblock In {\em Proceedings of the 4th USENIX conference on Networked systems
  design \& implementation\/} (2007), USENIX Association, pp.~20--20.

\bibitem{Google2018Failure}
Google cloud storage incident \#18003, Sept. 2018.
\newblock \url{https://status.cloud.google.com/incident/storage/18003}.

\bibitem{Gudmundsdottir19}
{\sc Gudmundsdottir, H.}
\newblock Systems@scale'19 talk: A tale of two performance analysis tools,
  2019.

\bibitem{guo2013failure}
{\sc Guo, Z., McDirmid, S., Yang, M., Zhuang, L., Zhang, P., Luo, Y., Bergan,
  T., Musuvathi, M., Zhang, Z., and Zhou, L.}
\newblock Failure recovery: when the cure is worse than the disease.
\newblock In {\em HotOS '13: Proceedings of the 14th Worksop on Hot Topics in
  Operating Systems\/} (2013).

\bibitem{Jaeger}
Jaeger: open-source, end-to-end distributed tracing.
\newblock \url{https://www.jaegertracing.io}.

\bibitem{Kaldor:2017gp}
{\sc Kaldor, J., Mace, J., Bejda, M., Gao, E., Kuropatwa, W., O'Neill, J., Ong,
  K.~W., Schaller, B., Shan, P., Viscomi, B., Venkataraman, V., Veeraraghavan,
  K., and Song, Y.~J.}
\newblock {Canopy: An end-to-end performance tracing and analysis system}.
\newblock In {\em SOSP '17: Proceedings of the 26th Symposium on Operating
  Systems Principles\/} (2017).

\bibitem{Lamport:1978tr}
{\sc Lamport, L.}
\newblock {Time, clocks, and the ordering of events in a distributed system}.
\newblock {\em Communications of the ACM 21}, 7 (July 1978).

\bibitem{Sifter}
{\sc Las-Casas, P., Papakerashvili, G., Anand, V., and Mace, J.}
\newblock Sifter: Scalable sampling for distributed traces, without feature
  engineering.
\newblock In {\em Proceedings of the ACM Symposium on Cloud Computing\/} (New
  York, NY, USA, 2019), SoCC ’19, Association for Computing Machinery,
  p.~312–324.

\bibitem{Las-Casas2019a}
{\sc Las-Casas, P., Papakerashvili, G., Anand, V., and Mace, J.}
\newblock Sifter: Scalable sampling for distributed traces, without feature
  engineering.
\newblock In {\em SoCC'19: Proceedings of the Ninth Symposium on Cloud
  Computing\/} (2019).

\bibitem{Lin:2014}
{\sc Lin, W., Xiao, X., and Ghinita, G.}
\newblock Large-scale frequent subgraph mining in mapreduce.
\newblock In {\em Data Engineering (ICDE), 2014 IEEE 30th International
  Conference on\/} (2014), IEEE, pp.~844--855.

\bibitem{Nagaraj:2012ur}
{\sc Nagaraj, K., Killian, C., and Neville, J.}
\newblock Structured comparative analysis of systems logs to diagnose
  performance problems.
\newblock In {\em NSDI '12: Proceedings of the 9\textsuperscript{th} USENIX
  conference on Networked Systems Design and Implementation\/} (2012).

\bibitem{Narayanan:2017:gv}
{\sc Narayanan, A., Chandramohan, M., Venkatesan, R., Chen, L., Liu, Y., and
  Jaiswal, S.}
\newblock graph2vec: Learning distributed representations of graphs.
\newblock {\em arXiv preprint arXiv:1707.05005\/} (2017).

\bibitem{Nijssen:2004}
{\sc Nijssen, S., and Kok, J.~N.}
\newblock Frequent graph mining and its application to molecular databases.
\newblock In {\em Systems, Man and Cybernetics, 2004 IEEE International
  Conference on\/} (2004), vol.~5, IEEE, pp.~4571--4577.

\bibitem{Odell:2017ei}
{\sc O'Dell, D.~H.}
\newblock {The Debugging Mindset}.
\newblock {\em ACM Queue 15}, 1 (Feb. 2017), 50.

\bibitem{Ohmann:2014cx}
{\sc Ohmann, T., Thai, K., Beschastnikh, I., and Brun, Y.}
\newblock {Mining precise performance-aware behavioral models from existing
  instrumentation}.
\newblock In {\em ICSE Companion '14: Companion Proceedings of the 36th
  International Conference on Software Engineering\/} (2014).

\bibitem{OpenStack}
Openstack web site.
\newblock \url{https://www.openstack.org}.

\bibitem{OpenTelemetry}
opentelemetry.
\newblock \url{https://opentelemetry.io}.

\bibitem{OpenTracing}
{{OpenTracing} website}.
\newblock \url{http://opentracing.io/}.

\bibitem{Iyer:2018td}
{\sc Padmanabha, A.~I., Liu, Z., and Jin, X.}
\newblock {ASAP}: Fast, approximate graph pattern mining at scale.
\newblock In {\em OSDI '18: Proceedings of the 13th USENIX Conference on
  Operating Systems Design and Implementation\/} (2018).

\bibitem{Reynolds:2006tl}
{\sc Reynolds, P., Killian, C., Wiener, J.~L., Mogul, J.~C., Shah, M., and
  Vahdat, A.}
\newblock Pip: detecting the unexpected in distributed systems.
\newblock In {\em NSDI '06: Proceedings of the 3\textsuperscript{rd} USENIX
  Symposium on Networked Systems Design and Implementation\/} (2006).

\bibitem{Sambasivan:2016bo}
{\sc Sambasivan, R.~R., Shafer, I., Mace, J., Sigelman, B.~H., Fonseca, R., and
  Ganger, G.~R.}
\newblock {Principled workflow-centric tracing of distributed systems}.
\newblock In {\em SoCC '16: Proceedings of the Seventh Symposium on Cloud
  Computing\/} (2016).

\bibitem{Sambasivan:2013tq}
{\sc Sambasivan, R.~R., Shafer, I., Mazurek, M.~L., and Ganger, G.~R.}
\newblock Visualizing request-flow comparison to aid performance diagnosis in
  distributed systems.
\newblock {\em IEEE Transactions on Visualization and Computer Graphics
  (Proceedings Information Visualization 2013) 19}, 12 (Dec. 2013).

\bibitem{Sambasivan:2011vw}
{\sc Sambasivan, R.~R., Zheng, A.~X., De~Rosa, M., Krevat, E., Whitman, S.,
  Stroucken, M., Wang, W., Xu, L., and Ganger, G.~R.}
\newblock {Diagnosing performance changes by comparing request flows}.
\newblock In {\em NSDI'11: Proceedings of the 8\textsuperscript{th} USENIX
  Conference on Networked Systems Design and Implementation\/} (2011).

\bibitem{Seno:2003}
{\sc Seno, M., Kuramochi, M., and Karypis, G.}
\newblock {PAFI: A Pattern Finding Toolkit}.
\newblock Tech. rep., ARMY HIGH PERFORMANCE COMPUTING RESEARCH CENTER
  MINNEAPOLIS MN, 2003.

\bibitem{Sigelman:2010uj}
{\sc Sigelman, B.~H., Barroso, L.~A., Burrows, M., Stephenson, P., Plakal, M.,
  Beaver, D., Jaspan, S., and Shanbhag, C.}
\newblock {Dapper, a large-scale distributed systems tracing infrastructure}.
\newblock Tech. Rep. dapper-2010-1, Google, Apr. 2010.

\bibitem{Loggly}
Solarwinds loggly.
\newblock https://www.loggly.com.

\bibitem{Amazon2011Failure}
Summary of the {Amazon} {EC2} and {Amazon} {RDS} service disruption in the {US}
  east region, Apr. 2011.
\newblock \url{http://aws.amazon.com/message/65648}.

\bibitem{Amazon2017Failure}
Summary of the {Amazon} {S3} service disruption in the {Nothern Virginia}
  {(US-EAST-1)} region, Feb. 2017.
\newblock \url{https://aws.amazon.com/message/41926/}.

\bibitem{Teixeira:2015bk}
{\sc Teixeira, C. H.~C., Fonseca, A.~J., Serafini, M., Siganos, G., Zaki,
  M.~J., and Aboulnaga, A.}
\newblock {Arabesque: a system for distributed graph mining}.
\newblock In {\em SOSP '15: Proceedings of the 25th Symposium on Operating
  Systems Principles\/} (2015).

\bibitem{HDFS}
The apache hadoop distributed file system, 2013.
\newblock \url{http://hadoop.apache.org/hdfs/}.

\bibitem{Weil:2006ti}
{\sc Weil, S.~A., Brandt, S.~A., Miller, E.~L., Long, D. D.~E., and Maltzahn,
  C.}
\newblock {Ceph}: a scalable, high-performance distributed file system.
\newblock In {\em OSDI '06: Proceedings of the 7\textsuperscript{th} USENIX
  Symposium on Operating Systems Design and Implementation\/} (2006).

\bibitem{Xu:2009ui}
{\sc Xu, W., Huang, L., Fox, A., Patterson, D., and Jordan, M.}
\newblock {Detecting large-scale system problems by mining console logs}.
\newblock In {\em SOSP '09: Proceedings of the 22\textsuperscript{nd} ACM
  Symposium on Operating Systems Principles\/} (2009).

\bibitem{Zhao:2017co}
{\sc Zhao, X., Rodrigues, K., Luo, Y., Stumm, M., Yuan, D., and Zhou, Y.}
\newblock Log20: Fully automated optimal placement of log printing statements
  under specified overhead threshold.
\newblock In {\em SOSP '17: Proceedings of the 26th Symposium on Operating
  Systems Principles\/} (2017).

\bibitem{Zhuang:2014vw}
{\sc Zhuang, Y., Gessiou, E., Portzer, S., Fund, F., Muhammad, M.,
  Beschastnikh, I., and Cappos, J.}
\newblock Netcheck: network diagnoses from blackbox traces.
\newblock In {\em NSDI'14: Proceedings of the 11th USENIX Conference on
  Networked Systems Design and Implementation\/} (Apr. 2014), USENIX
  Association.

\end{thebibliography}
}

\end{document}